\documentclass[aps,twocolumn,groupedaddress,superscriptaddress,nofootinbib,floatfix]{revtex4-1}
\usepackage{graphicx}
\usepackage{color}
\usepackage{pifont}
\usepackage{amsmath,amsfonts,amssymb}

\usepackage{bbold}
\usepackage{t1enc}
\usepackage{latexsym} 
\usepackage{cancel}

\DeclareMathSymbol{\shortminus}{\mathbin}{AMSa}{"39}
\DeclareMathSymbol{\shm}{\mathbin}{AMSa}{"39}

\newcommand{\met}{\cancel{E}_T}

\begin{document}

\vspace*{-2.1cm}
\begin{flushright}
IFT-UAM/CSIC-24-112 \\
CERN-TH-2024-123
\vspace*{-0.2cm}
\end{flushright}
\vspace{0.cm}

\vspace{-2cm}
\begin{flushright}
\end{flushright}
\vspace{1cm}

\title{Toponium Hunter's Guide}

\author{J. A. Aguilar-Saavedra}
\affiliation{Instituto de F\'\i sica Te\'orica, IFT-UAM/CSIC, c/ Nicolás Cabrera 13-15, 28049 Madrid, Spain}

\begin{abstract}
We address the discovery and characterisation of toponium production at the Large Hadron Collider. In the dilepton decay mode, multivariate analyses of spin and colour observables could provide evidence that an excess of events present near the $t \bar t$ threshold corresponds to a spin-zero colour singlet. The semileptonic decay mode may also exhibit an excess near threshold, but is not expected to play any role in the toponium characterisation.
\end{abstract}

\maketitle

\section{Introduction}

Measurements of quantum entanglement between the spins of top quark pairs produced at the Large Hadron Collider (LHC) have recently been performed by the ATLAS~\cite{ATLAS:2023fsd} and CMS~\cite{CMS:2024pts} Collaborations, in a kinematical region near the $t \bar t$ production threshold. The ATLAS measurement exhibits a quite sizeable discrepancy with respect to next-to-leading order (NLO) predictions of the Standard Model (SM). On the other hand, the CMS collaboration has found that, remarkably, data is very well explained provided the SM NLO prediction is supplemented by the production of {\em toponium}, a yet unobserved $t \bar t$ bound state of extremely short lifetime. Bound states of $t \bar t$ are a prediction of the SM, and are essentially a non-perturbative phenomenon. In hadron collisions, the formation of a $J^P = 0^-$ colour-singlet resonance is predicted~\cite{Fadin:1990wx,Hagiwara:2008df,Kiyo:2008bv,Sumino:2010bv}, while in $e^+ e^-$ collisions a $J^P = 1^-$ resonance is expected near threshold~\cite{Fadin:1987wz,Strassler:1990nw,Sumino:1992ai}. 

The renewed interest in toponium production motivates a thorough investigation of strategies towards its potential discovery at the LHC. At hadron colliders the presence of a toponium `signal' may be spotted by deviations with respect to the predictions of perturbative QCD near the $t \bar t$ threshold. These deviations comprise:
\begin{enumerate}
\item[(i)] an event excess;
\item[(ii)] differences in observables characterising the spin of the $t \bar t$ pair, which arise because $t \bar t$ from toponium decay areproduced in a spin-singlet state;
\item[(iii)] differences in observables characterising the $t \bar t$ colour connection: for toponium decay the $t \bar t$ pair is a colour-singlet, while $t \bar t$ in the continuum is dominated by the colour-octet state.\footnote{For brevity we will often refer to $t \bar t$ production in perturbative QCD as production in the continuum, as opposed to resonant toponium production.}
\end{enumerate}
Previous work~\cite{Fuks:2021xje} has addressed the prospects to discover toponium in the dilepton decay channel  $t \bar t \to \ell \nu b \ell \nu \bar b$, with $\ell = e,\mu$, as an excess of events near threshold. This excess can be enhanced via suitable kinematical cuts on the dilepton invariant mass $m_{\ell \ell}$ and the laboratory-frame azimuthal angle difference $\Delta \phi_{\ell \ell}$. These two variables are related to the $t \bar t$ spin correlation: the $t \bar t$ spin-singlet state tends to produce closer leptons, which also have smaller invariant mass. However, that simple strategy does not seem sufficient to provide strong evidence for toponium production. It also has the disadvantage that $\Delta \phi$ is quite sensitive to boosts of the $t \bar t$ pair in the transverse plane.\footnote{The $\Delta \phi_{\ell \ell}$ distribution is known to exhibit a mismodeling~\cite{ATLAS:2019zrq,CMS:2019nrx} that cannot be attributed to the toponium contribution, since it is present far above threshold. Next-to-next-to-leading order corrections~\cite{Behring:2019iiv} improve the agreement, but still data and predictions exhibit some discrepancies~\cite{topLHCwg}.} Lepton angles defined in the rest frame of the parent top quark are experimentally more challenging but also more robust from the theoretical point of view.

The discovery and characterisation of toponium is, admittedly, a formidable task that requires a very accurate theoretical modeling. As a first step, it is useful to investigate the observables that would reveal its properties, thereby allowing to characterise an excess near threshold as `toponium'. This is the purpose of this work. Our goal is not to give precise sensitivity estimations: these would not be realistic in the absence of systematic uncertainties, which can only be evaluated by the experiments. Instead, our focus is on the strategy that experiments could follow, including a comparative study of the observables that might be used for that characterisation. The strategy proposed is, in any case, well motivated by the estimations we provide of the statistical sensitivity to observe deviations.

The main focus of this work is the dilepton decay mode. Provided the systematic uncertainties are under control, significant deviations with respect to the predictions of perturbative QCD could be measured near threshold, in the total number of events, and spin / colour observables. For the analysis of the latter two, a multivariate analysis could be very useful, as we will show. 
The semileptonic decay mode of the top quark pair, which has a larger branching ratio than the dilepton one, may also be useful to spot an event excess near threshold, but not for the characterisation of the toponium properties. These points are further elaborated in an appendix.

\section{Event generation, selection and reconstruction}
\label{sec:2}

The production of $t \bar t$ in the continuum is modeled with the SM process $pp \to b \bar b WW$, which includes $t \bar t$ as well as non-resonant diagrams. It is is generated with {\scshape MadGraph}~\cite{Alwall:2014hca} at the leading order, using NNPDF 3.1~\cite{NNPDF:2017mvq} parton density functions and setting as factorisation and renormalisation scale half the  transverse mass, $Q = 1/2 \sum_i (m_i^2 + p_{T i}^2)^{1/2}$, with $p_T$ the transverse momentum in the usual notation and $i$ labelling the different particles. We set the top mass to $m_t = 172.5$ GeV. Toponium production is modeled as a pseudo-scalar resonance $\eta_t$ with mass $m_{\eta_t} \simeq 2 m_t - 2 = 343$ GeV, width $\Gamma_{\eta_t} \simeq 2 \Gamma_t = 3$ GeV~\cite{Maltoni:2024csn}, and interactions
\begin{equation}
\mathcal{L} = - g_{gg\eta_t} G_{\mu \nu}^a \tilde G^{\mu \nu a} \, \eta_t - i g_{tt\eta_t} \bar t \gamma_5 t \, \eta_t
\end{equation}
The effective $gg\eta_t$ interaction stands for the triangle loop diagram with a top quark. For our study this is a good approximation because being a spin-zero particle, the only feature required from the production is the cross section, which is fitted from calculations of $t \bar t$ production near threshold with non-relativistic effects~\cite{Hagiwara:2008df,Kiyo:2008bv,Sumino:2010bv}. The spatial size of toponium also gives rise to differences in top momentum distrubutions, which can be implemented via a Green function reweighting~\cite{Fuks:2021xje}. However, these differences are rather small, c.f. figs. 13, 14 from Ref.~\cite{Sumino:2010bv}, and not relevant for our analysis.

We generate high-statistics samples with $3 \times 10^7$ events for $pp \to b \bar b W^+ W^- \to b \bar b \ell^+ \nu \ell^- \nu$, with $\ell = e,\mu$, and $3 \times 10^6$ events for $pp \to \eta_t \to b \bar bW^+ W^- \to b \bar b \ell^+ \nu \ell^- \nu$, with $\ell = e,\mu$.
Hadronisation and parton showering is performed with {\scshape Pythia}~\cite{Sjostrand:2007gs} and detector simulation with {\scshape Delphes}~\cite{deFavereau:2013fsa} using the default card for the CMS detector. Jets are reconstructed with {\scshape FastJet}~\cite{Cacciari:2011ma}  using the anti-$k_T$ algorithm~\cite{Cacciari:2008gp} with radius $R=0.4$. A probabilistic $b$-tagging is applied corresponding to the 70\% efficiency working point~\cite{CMS:2012feb}.
We apply the kinematical selection criteria of the CMS entanglement measurement~\cite{CMS:2024pts}, which are inherited from the spin correlation measurement~\cite{CMS:2019nrx}:
\begin{itemize}
\item Two opposite-sign charged leptons with pseudo-rapidity $|\eta| \leq 2.4$, the leading one with transverse momentum $p_T \geq 25$ GeV, and the trailing one with $p_T \geq 20$ GeV. Their invariant mass $m_{\ell \ell}$ has to be larger than 20 GeV.
\item Two jets with $|\eta| \leq 2.4$ and $p_T \geq 30$ GeV, one of them $b$-tagged.
\item When the two charged leptons have the same flavour, the invariant mass window $76 \leq m_{\ell \ell} \leq 106$ GeV is excluded, and a lower cut $\met \geq 40$ GeV is placed on the missing transverse energy (MET). 
\end{itemize}
The overall efficiency of this event selection is 0.14.

The final state is reconstructed assuming the kinematics of nearly-on-shell production of a top quark pair that decays $t \bar t \to b W^+ \bar b W^- \to b \ell^+ \nu \bar b  \ell^- \nu$. Furthermore, it is assumed that the MET originates from the two escaping neutrinos. A minimisation is performed to find the values of the neutrino momenta that are most compatible with the assumed $t \bar t \to b W \bar b W$ kinematics. For given neutrino momenta $p_{\nu_1}$, $p_{\nu_2}$ (three unknowns for each momentum), 
the $W$ and top quark momenta are reconstructed as
\begin{align}
& p_{W_1} = p_{\ell_1} + p_{\nu_1}  \,, \quad p_{t_1} = p_{\ell_1} + p_{\nu_1} + p_{b_1} \,, \notag \\
& p_{W_2} = p_{\ell_2} + p_{\nu_2}  \,, \quad p_{t_2} = p_{\ell_2} + p_{\nu_2} + p_{b_2} \,,
\end{align}
and their reconstructed invariant masses are labelled as $m_{W_{1,2}}$, $m_{t_{1,2}}$.
In events with two $b$-tagged jets we select them for the reconstruction, and in events with only one $b$-tagged jet we attempt the reconstruction selecting also one of the two untagged jets with largest $p_T$.
The labelling of the two neutrinos $\nu_1$, $\nu_2$ is defined by the accompanying charged lepton, while there are two possible pairings of charged leptons and $b$ quarks to reconstruct the top quarks. For each pairing, the neutrino momenta are chosen as the ones that minimise the quantity
\begin{widetext}
\begin{eqnarray}
\chi^2 & = & \frac{(m_{W_1}- M_W)^2}{\sigma_W^2} + \frac{(m_{W_2}- M_W)^2}{\sigma_W^2}
+ \frac{(m_{t_1} - m_t)^2}{\sigma_t^2} + \frac{(m_{t_2}- m_t)^2}{\sigma_t^2} \notag \\
& & + \frac{\left[ (p_{\nu_1})_x + (p_{\nu_2})_x - (\met)_x \right]^2}{\sigma_p^2} 
+ \frac{\left[ (p_{\nu_1})_y + (p_{\nu_2})_y - (\met)_y \right]^2}{\sigma_p^2} \notag \\
& & + \frac{\left[ (p_{t_1})_T - (p_{t_2})_T \right]^2}{\sigma_p^2} \,,
\label{ec:chi}
\end{eqnarray}
\end{widetext}
and among the two possible pairings between charged leptons and $b$ quarks, we select the pairing with smallest $\chi^2$.
The terms in the first line favour solutions where the reconstructed masses are close to the true ones, taken as $M_W = 80.4$ GeV, $m_t = 172.5$ GeV, but without explicitly requiring that any of the particles is on its mass shell. The denominators $\sigma_W = 7.5$ GeV, $\sigma_t = 11.5$ GeV~\cite{Aguilar-Saavedra:2021ngj} represent typical experimental resolutions for the reconstructed width of the top quarks and $W$ bosons.
The terms in the second line account for our assumption that MET results from the two neutrinos, while also considering potential mismeasurements. The term in the third line avoids solutions with large transverse momentum imbalance for the two top quarks, with $\sigma_p = 20$ GeV~\cite{Aguilar-Saavedra:2021ngj} a typical value of the experimental resolution for top transverse momenta. Variations of these values basically give the same results.

\begin{figure}[t]
\begin{center}
\begin{tabular}{c}
\includegraphics[width=8.4cm,clip=]{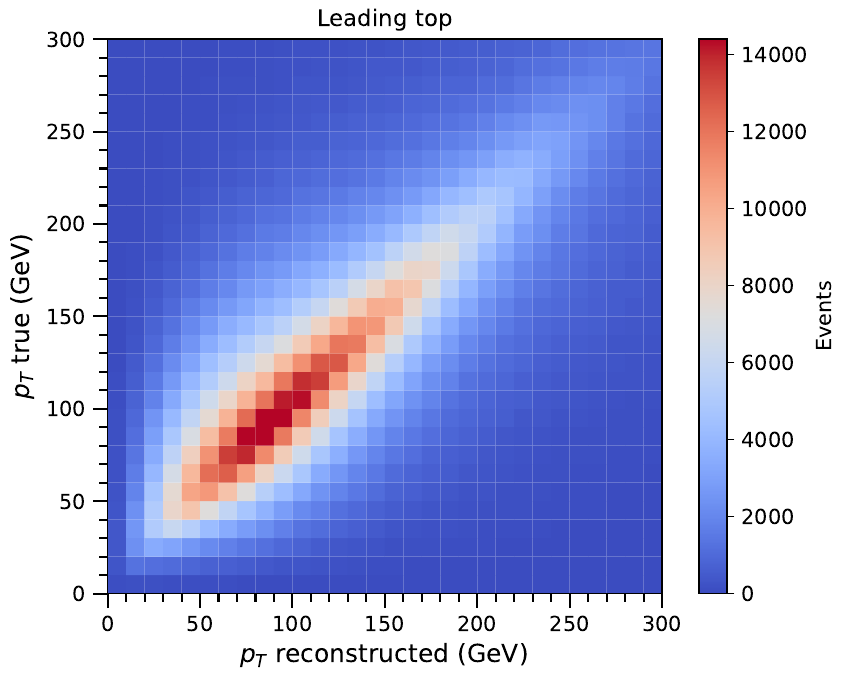} \\
\includegraphics[width=8.4cm,clip=]{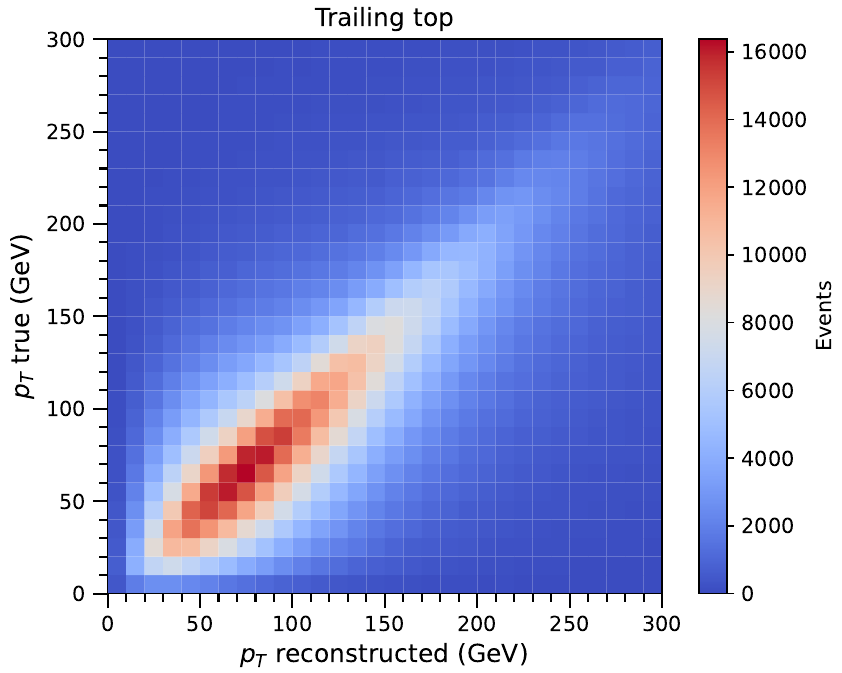}
\end{tabular}
\end{center}
\caption{Reconstructed versus true transverse momenta of the top quarks, for the $t \bar t$ process.}
\label{fig:reco2}
\end{figure}

\begin{figure}[htb]
\begin{center}
\begin{tabular}{c}
\includegraphics[width=8.4cm,clip=]{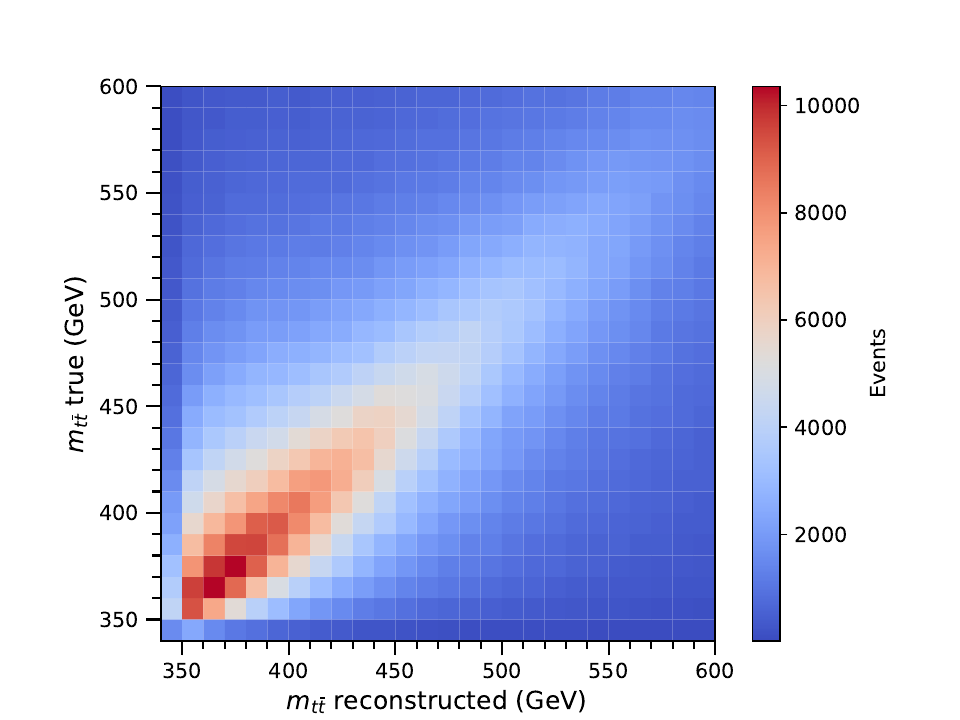} \\
\includegraphics[width=8.4cm,clip=]{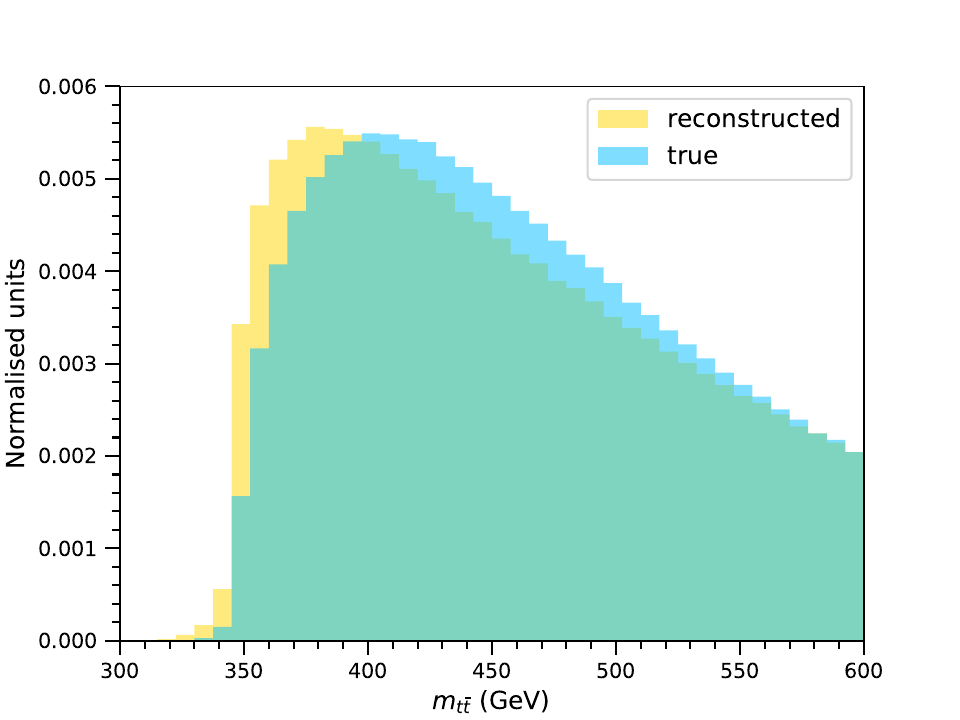}
\end{tabular}
\end{center}
\caption{Top: reconstructed versus true $t \bar t$ invariant mass for the $t \bar t$ process. Bottom: one-dimensional projection.}
\label{fig:reco3}
\end{figure}

Radiation may result in energy loss for $b$ quark jets. To address this, our minimisation process includes the possibility of scaling the $b$ quark four-momenta to higher values, up to one standard deviation of the expected jet energy resolution of 15\%~\cite{CMS:2016lmd}. This results in a total number of 8 variables for the minimisation. 
The stability of the solution for the minimsation problem is verified by starting with three different initial values: (a) all the missing energy corresponds to $\nu_1$, and $(p_{\nu_2})_z = 0$; (b) the same with $\nu_1 \leftrightarrow \nu_2$; (c) the missing energy is equally shared by the two neutrinos. Although there are small numerical differences in the minima found, the resulting kinematical distributions are nearly identical. 

\begin{figure}[htb]
\begin{center}
\begin{tabular}{c}
\includegraphics[width=8.4cm,clip=]{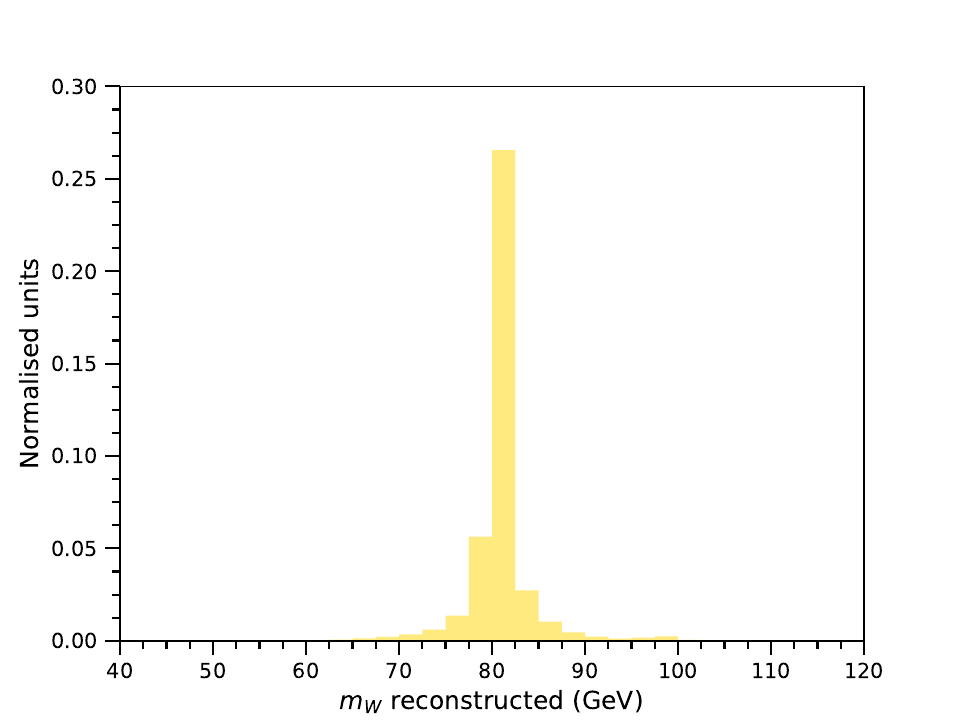} \\
\includegraphics[width=8.4cm,clip=]{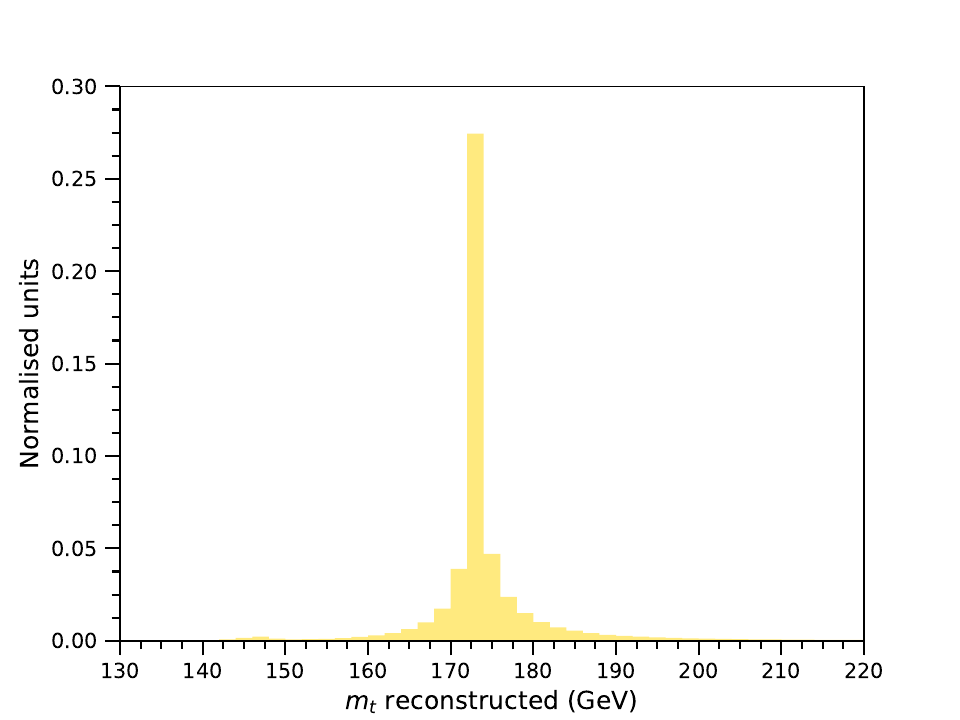}
\end{tabular}
\end{center}
\caption{Reconstructed $W$ boson and top quark masses, for the $t \bar t$ process.}
\label{fig:reco1b}
\end{figure}

Figures~\ref{fig:reco2}--\ref{fig:reco1b} show the performance of the reconstruction for $p p \to t \bar t \to b W^+ \bar b W^- \to b \ell^+ \nu \bar b \ell^- \nu$ events, with two nearly-on-shell top quarks. In fig.~\ref{fig:reco2} we present two-dimensional plots for the reconstructed $p_T$ versus true $p_T$ for the leading and trailing top quark. In fig.~\ref{fig:reco3} we present the true versus reconstructed invariant mass $m_{t \bar t}$. We observe a small shift of the reconstructed versus the true value; however, we do not attempt to correct for it because we address the observability of toponium at the detector level. Figure~\ref{fig:reco1b} shows the reconstructed $W$ boson and top quark masses.

\begin{figure}[htb]
\begin{center}
\begin{tabular}{c}
\includegraphics[width=8.4cm,clip=]{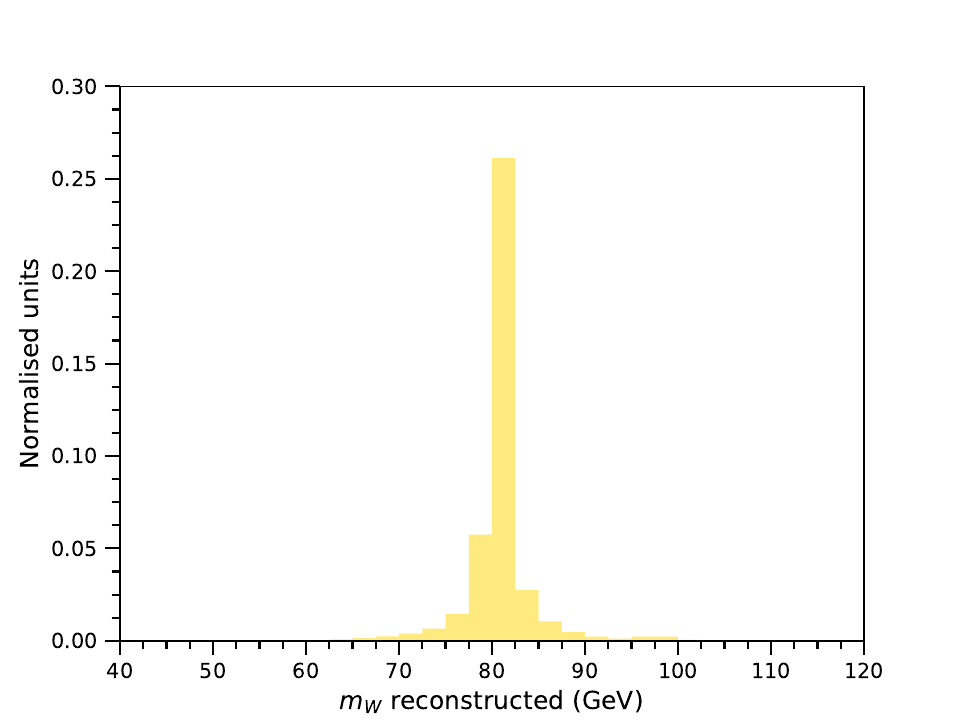} \\
\includegraphics[width=8.4cm,clip=]{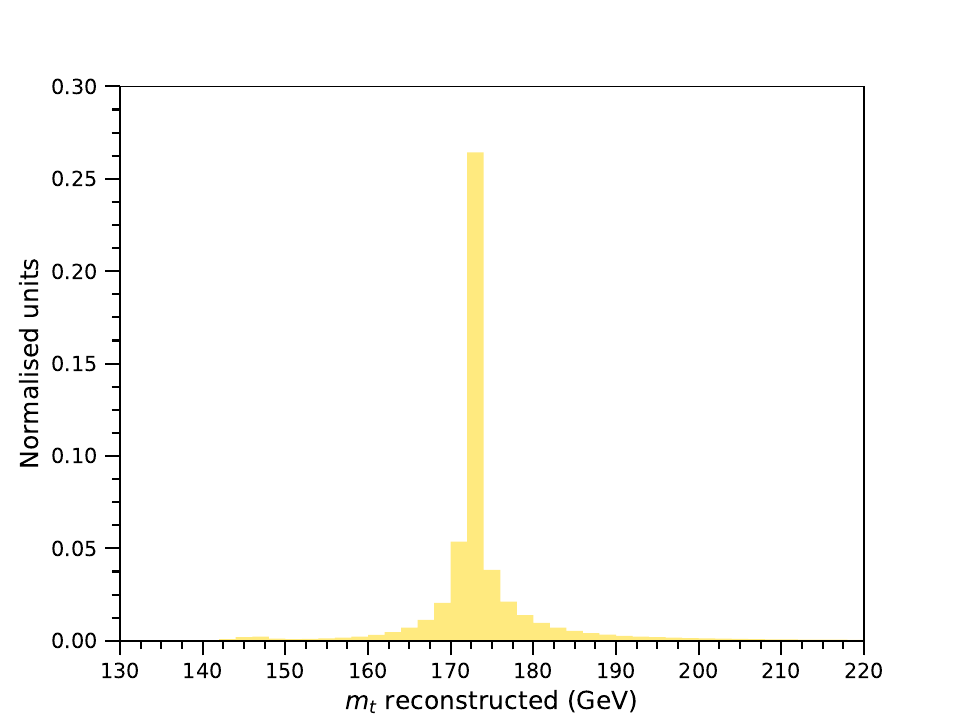}
\end{tabular}
\end{center}
\caption{Reconstructed $W$ boson and top quark masses, for the $b \bar b WW$ process.}
\label{fig:reco1}
\end{figure}

This $t \bar t$ reconstruction is applied to $b \bar b WW$ and toponium events. The $W$ boson and top quark reconstructed masses for $b \bar b WW$ are presented in fig.~\ref{fig:reco1}. Although this process has diagrams that correspond to single top production, we observe that events are quite compatible with a $t \bar t$ kinematics (compare with fig.~\ref{fig:reco1b}). Therefore, we do not apply any `quality' cut on the reconstruction, but keep the whole samples fulfilling the selection criteria.

\section{Multivariate analysis for toponium characterisation}

As previously discussed, the presence of a toponium `signal' may be verified by an event excess, and deviations in observables characterising the spin and colour of the $t \bar t$ pair. While the event excess can be spotted by a simple counting, deviations in spin and colour observables require a more sophisticated analysis. To this end, we use an unbinned multivariate test as proposed in Ref.~\cite{Williams:2010vh} that proceeds in two steps. First, a neural network (NN) is trained to discriminate between `signal' (toponium) and `background' ($b \bar b WW$) events, using variables that we describe in the following subsections. Subsequently, the NN score evaluated on signal plus background, and background-only samples, provides two one-dimensional distributions to which a Kolmogorov-Smirnov test~\cite{K,S} can be applied to determine the statistical compatibility of both. Using this approach in an experimental analysis, the NNs would be trained with Monte Carlo simulation, and the background-only sample test would be generated with Monte Carlo too. Other statistical methods~\cite{Grosso:2023scl} could be used as well, which skip the need of NN training on signal and background. Preliminary tests show that the expected statistical sensitivity is similar.\footnote{I thank Gaia Grosso for performing some tests using the NPLM method.}

For the training and testing of each NN a standardisation of the inputs, based on the background distributions, is performed. Each NN is trained with two samples of $1.5 \times 10^4$ signal and $1.5 \times 10^4$ background events.
The architecture of the NNs is not crucial for the discrimination. We use NNs with two hidden layers of 1024 and 128 nodes, with Rectified Linear Unit (ReLU) activation for the hidden layers and a sigmoid function for the output one. For simplicity we keep the same NN architecture even if the number of inputs (5 for spin, 38 for colour) is very diffferent. We have tested that other NN architectures provide similar results.
The NN optimisation relies on the binary cross-entropy loss function, using the Adam~\cite{Adam} optimiser. Overtraining is avoided by monitoring the NN performance on validation samples, of the same size as the training samples, and stopping the training when the performance ceases to improve. The NNs are implemented using {\scshape Keras}~\cite{keras} with a {\scshape TensorFlow} backend~\cite{tensorflow}.

\subsection{Spin variables}
\label{sec:3.1}

\begin{figure}[t]
\begin{center}
\includegraphics[width=8.4cm,clip=]{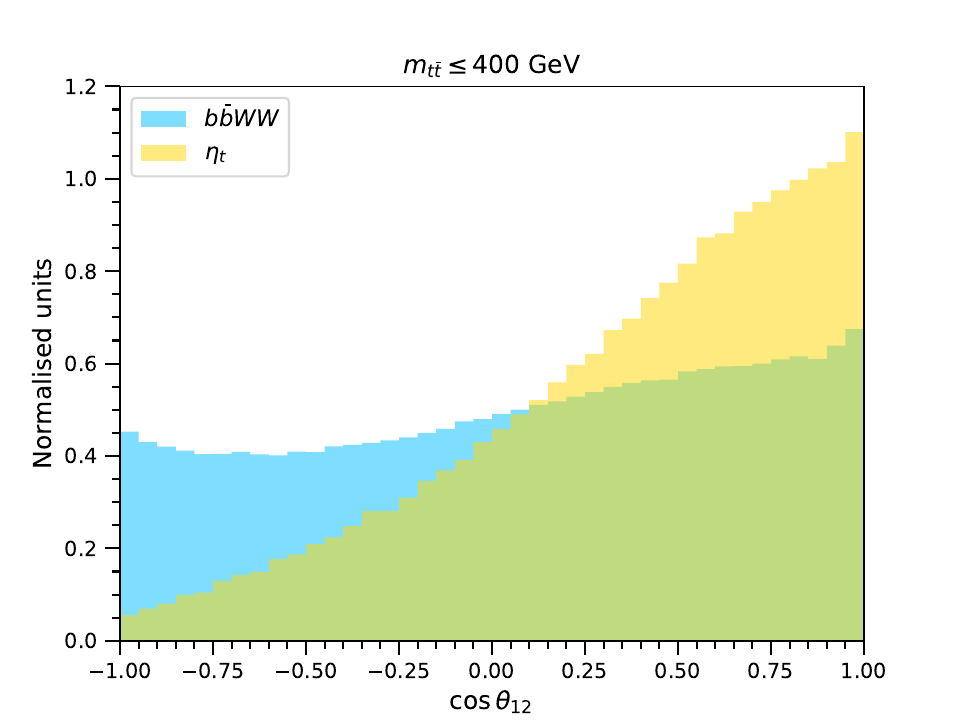}
\end{center}
\caption{Kinematical distribution of $\cos \theta_{12}$ for $b \bar b WW$ and the toponium signal.}
\label{fig:cosD}
\end{figure}

The spin state of the $t \bar t$ pair is fully characterised by the four-dimensional distribution of polar and azimuthal angles of the two charged leptons in the rest frame of their parent top quark. As reference system we use the so-called helicity basis~\cite{Bernreuther:2015yna} $(\hat r, \hat n, \hat k)$, with the axes defined as
\begin{itemize}
\item K-axis (helicity): $\hat k$ is a normalised vector in the direction of the top quark three-momentum in the $t \bar t$ rest frame.
\item R-axis: $\hat r$ is in the production plane and defined as $\hat r = (\hat p_p - \cos \theta_t \hat k)/\sin \theta_t$, with $\hat p_p = (0,0,1)$ the momentum of one of the initial protons and $\theta_t$ the production angle in the centre-of-mass (c.m.) frame.
\item N-axis: $\hat n = \hat k \times \hat r$ is orthogonal to the production plane.
\end{itemize}
The same basis is used for the top quark and anti-quark, and with respect to this basis, the rest-frame angles $(\theta_1,\phi_1)$ for $\ell^+$ and $(\theta_2,\phi_2)$ for $\ell^-$ are defined. From the kinematical distribution of these four quantities, the spin density operator of the $t \bar t$ pair can be determined. However, close to threshold where the top quarks are nearly at rest in the $t \bar t$ rest frame, an accurate determination of the helicity basis is difficult. Therefore, in addition to $\theta_{1,2}$ and $\phi_{1,2}$ we include in our set the angle between the two lepton momenta $\theta_{12}$, which is basis-independent. We present this distribution in fig.~\ref{fig:cosD}, for events with reconstructed invariant mass $m_{t \bar t} \leq 400$ GeV.

\subsection{Colour variables}
\label{sec:3.2}

\begin{figure}[t]
\begin{center}
\begin{tabular}{c}
\includegraphics[width=8.4cm,clip=]{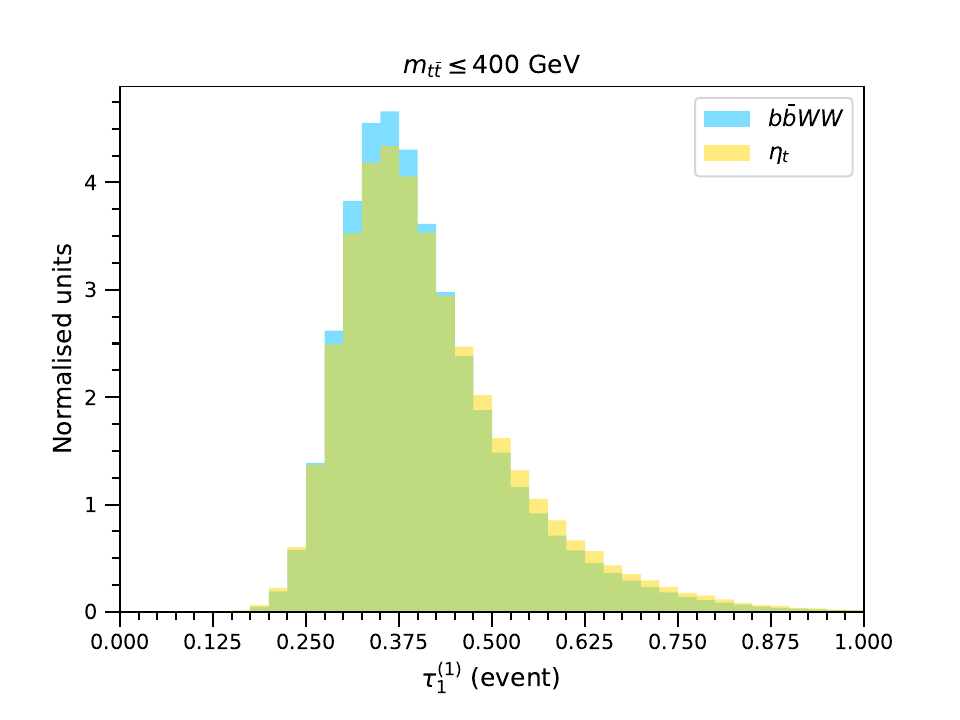} \\
\includegraphics[width=8.4cm,clip=]{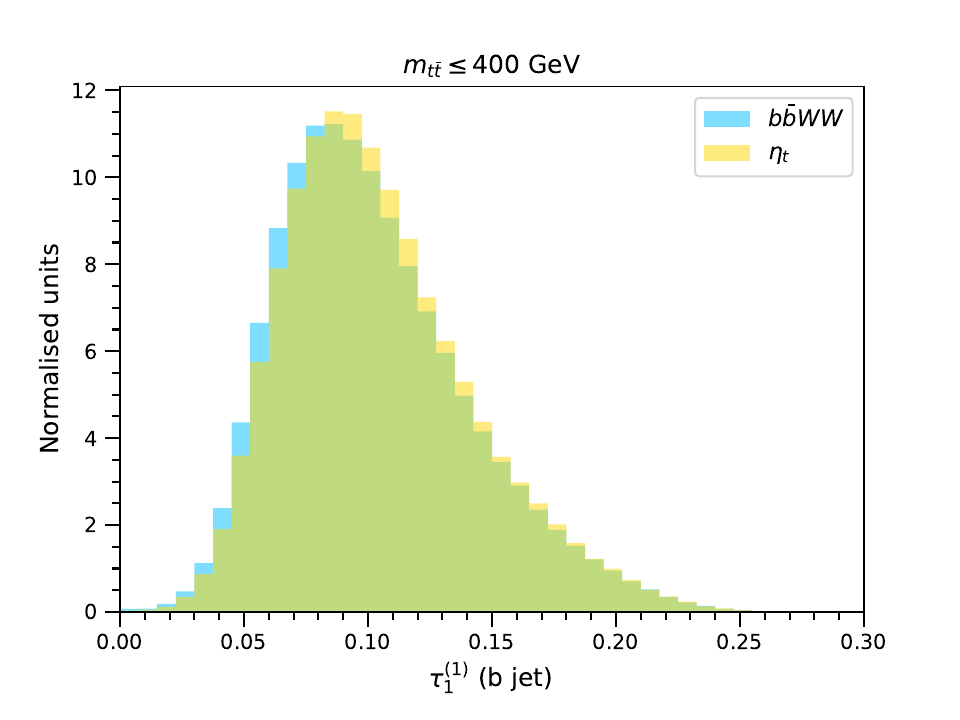}
\end{tabular}
\end{center}
\caption{$\tau_1^{(1)}$ of the event (top) and of the $b$-tagged jet (bottom), for $b \bar b WW$ and the toponium signal.}
\label{fig:col}
\end{figure}

The discrimination between $b \bar b WW$ and $\eta_t$ is mainly based on the global shape of the event, which is parameterised by a set of variables adapted from the ones introduced to characterise jet substructure~\cite{Thaler:2010tr,Thaler:2011gf,Datta:2017rhs}. We define them as
\begin{equation}
\tau_n^{(\beta)} = \frac{1}{\sum_i E_i} \sum_i p_{T i} \operatorname{min} \left\{ \Delta R_{1i}^\beta ,  \Delta R_{2i}^\beta , \dots  \Delta R_{ni}^\beta \right\} \,,
\label{ec:tau}
\end{equation}
with $i$ labelling the detector-level particles in the event, and $p_{Ti}$, $E_i$ their transverse momentum and energy, repectively; $\Delta R_{k i} = \left[ \Delta \eta_{ki}^2  + \Delta \phi_{ki}^2 \right]^{1/2}$ is the lego-plot distance between the momentum of the particle $i$ and the axis $k=1,\dots,n$. In contrast to their widespread use for jet substructure, here we use $\tau_n^{(\beta)}$ to characterise the global structure of the event, with $n=1,\dots,6$ and $\beta = 1,2,0.5$.  Note that the variables in (\ref{ec:tau}) are not equivalent to the $N$-jettiness introduced in Ref.~\cite{Banfi:2010pa}. 
Additionally, we use the subjettiness of the two jets that are identified by the kinematical reconstruction (c.f. section~\ref{sec:2}) as corresponding to the hadronisation of the two $b$ quarks. In this case, we use $n=1,\dots,3$ and $\beta = 1,2,0.5$. The kinematical distributions of $\tau_n^{(\beta)}$ are not very different for  $b \bar b WW$ and toponium, but altogether these sets of variables provide a good discrimination between the colour-singlet and colour-octet configurations.
For illustration, fig.~\ref{fig:col} shows $\tau_1^{(1)}$ of the event and the $b$-tagged jet, for $b\bar bWW$ and the toponium signal. We have verified that including higher-order $\tau_n^{(\beta)}$ does not improve the results.

Additional variables providing some discrimination power are the jet multiplicity (for which we consider $p_T \geq 20$ GeV), and  the total number of detector-level particles in the event.  The jet pull~\cite{Gallicchio:2010sw} is an observable designed to probe colour flow between jets, which has been used by the D0~\cite{D0:2011lzz} and ATLAS~\cite{ATLAS:2015ytt} Collaborations to measure colour flow in the hadronic decays of $W$ bosons produced in the semileptonic decay of $t \bar t$ pairs. However, we find that these variables are strongly dependent on $m_{t \bar t}$, and bin migrations wash out the differences between $b \bar b WW$ and toponium, rendering their kinematical distributions nearly identical at the reconstructed level.

\section{Statistical sensitivity to toponium}

The toponium signal may be spotted by its effect near the $t \bar t$ threshold, which motivates an analysis in different bins of $m_{t \bar t}$. The bin size is a compromise between sample size and signal to background ratio. The Kolmogorov-Smirnov test used to detect deviations is more powerful for larger samples but, on the other hand, the signal (toponium) to background ($b \bar b WW$) ratio is larger near threshold. Additionally, the energy resolution broadens the toponium signal to higher values of $m_{t \bar t}$. We therefore use bins of 20 GeV, with the first bin $m_{t \bar t} \leq 360$ GeV. 

For the toponium signal, Ref.~\cite{Fuks:2021xje} used calculations in~\cite{Sumino:2010bv} to extract a toponium cross section $\sigma(\eta_t) = 6.43$ pb at 13 TeV. On the other hand, Ref.~\cite{Maltoni:2024csn} used calculations in~\cite{Kiyo:2008bv} to extract a toponium cross section $\sigma(\eta_t) = 3.6$ pb at 14 TeV (at 13 TeV the cross section is 15\% smaller). We use the former as our baseline benchmark, since the entanglement measurement performed by the CMS Collaboration~\cite{CMS:2024pts} obtained very good agreement with data using that value, and provide alternative results for the lower cross section.
For the $b \bar b WW$ background there are no calculations of the cross section beyond the leading order (LO); at LO its cross section is 1.1 times larger than for $t \bar t$. At NLO the $t \bar t$ cross section using NNPDF 3.1 parton density functions is 671 pb. Then, we assume a next-to-leading order cross section of $k \times 671$ pb for $b \bar b WW$, with $k = 1.1$ our baseline choice. We also explore $k = 1.5$ as conservative background estimate. We note that the $t \bar t$ cross section is known to next-to-next-to-leading order~\cite{Czakon:2013goa}. However, since the predictions in Refs.~\cite{Kiyo:2008bv,Sumino:2010bv} from which the toponium cross sections are extracted are at NLO in perturbative QCD, for consistency we also use background predictions at NLO. For the baseline benchmark the expected number of events in each $m_{t \bar t}$ bin is collected in Table~\ref{tab:SB}, for luminosities corresponding to LHC Run 2 (140 fb$^{-1}$) and Run 2$+$3 (350 fb$^{-1}$). We also include the signal ($S$) to background ($B$) ratio in the last column.

\begin{table}[htb]
\begin{center}
\begin{tabular}{cccccc}
& \multicolumn{2}{c}{Run 2} & \multicolumn{2}{c}{Run 2$+$3} \\ 
$m_{t \bar t}$ (GeV) & $\eta_t$ & $b \bar b WW$ & $\eta_t$ & $b \bar b WW$ & $S/B$ \\
$\leq 360$ & 1640 & 38160 & 4370 & 101680 & $0.041$  \\    
$360-380$ & 1180 & 57000 & 3140 & 151890 & $0.020$ \\
$380-400$ & 740 & 58000 & 1970 & 154540 & $0.013$ \\
$400-420$ & 500 & 55160 & 1340 & 146950 & $0.009$ \\
\end{tabular}
\caption{Expected number of toponium and $b \bar b WW$ events in selected bins of $m_{t \bar t}$, for the baseline benchmark.}
\label{tab:SB}
\end{center}
\end{table}

\begin{figure}[htb]
\begin{center}
\includegraphics[width=8.4cm,clip=]{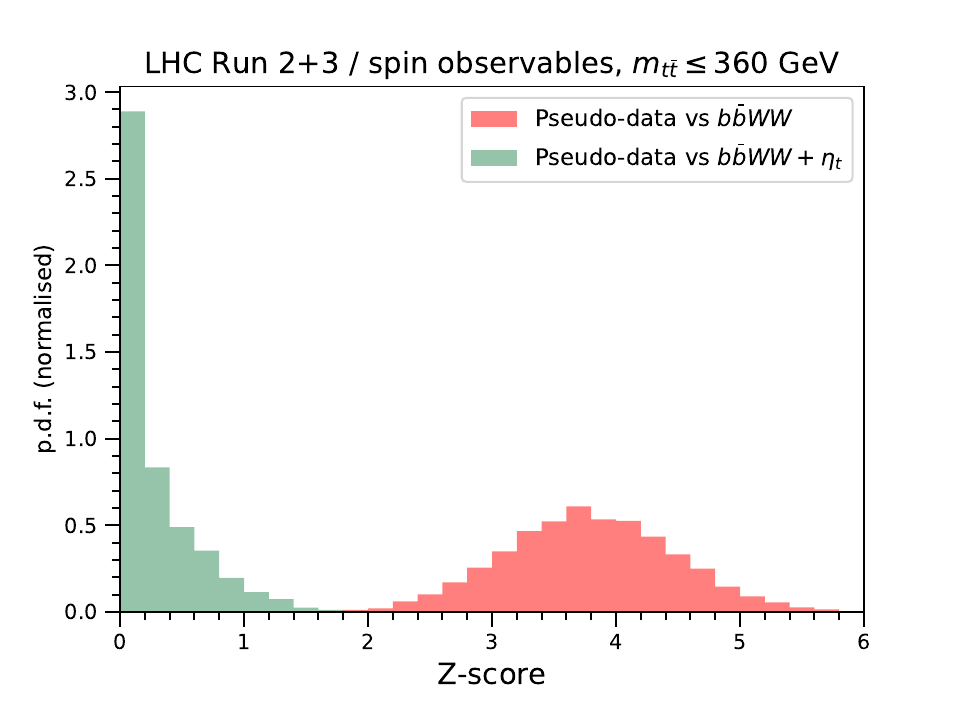}
\end{center}
\caption{Example of distributions for the statistical significance of the deviations between pseudo-data (with injected toponium) and the $b \bar b WW$ and $b \bar b WW+ \eta_t$ hypotheses, obtained from pseudo-experiments. }
\label{fig:Zscore}
\end{figure}

\begin{figure*}[htb]
\begin{center}
\begin{tabular}{cc}
\includegraphics[width=8.4cm,clip=]{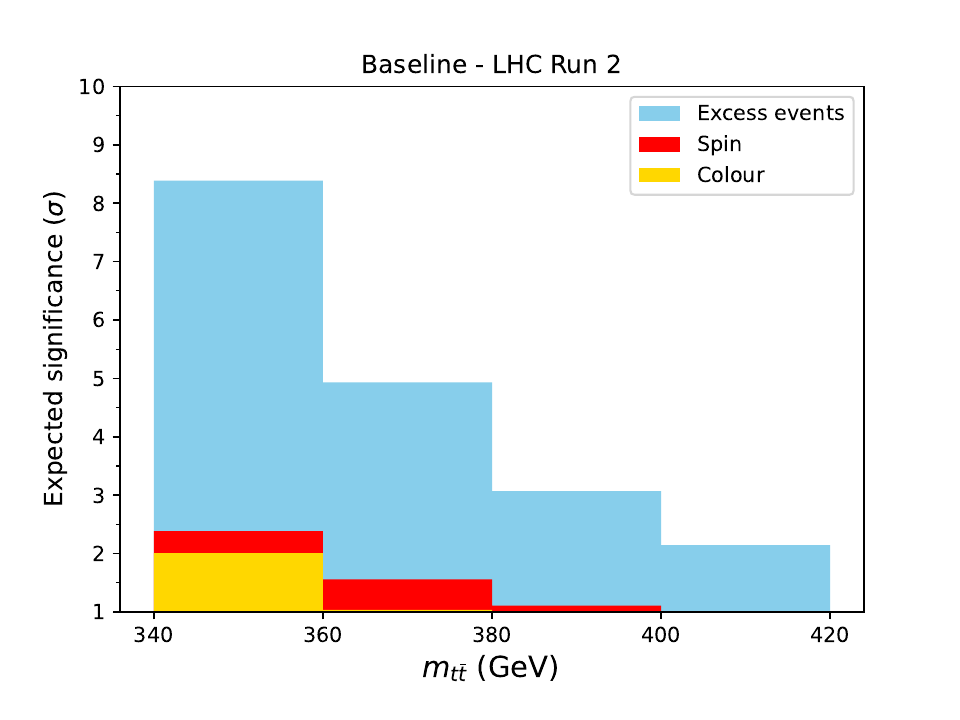} &
\includegraphics[width=8.4cm,clip=]{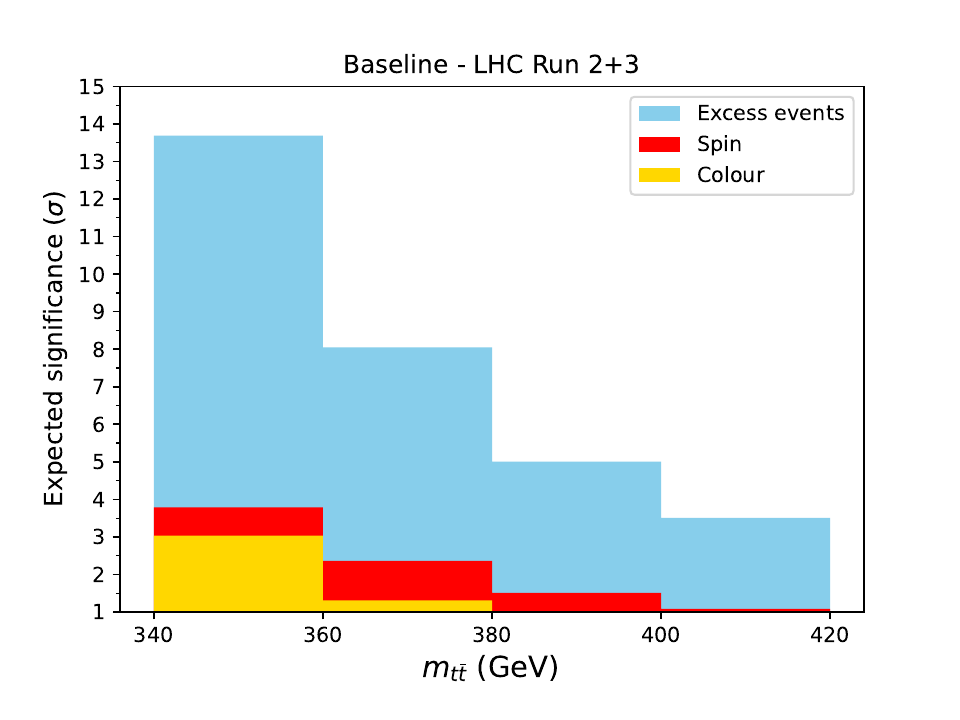} \\
\includegraphics[width=8.4cm,clip=]{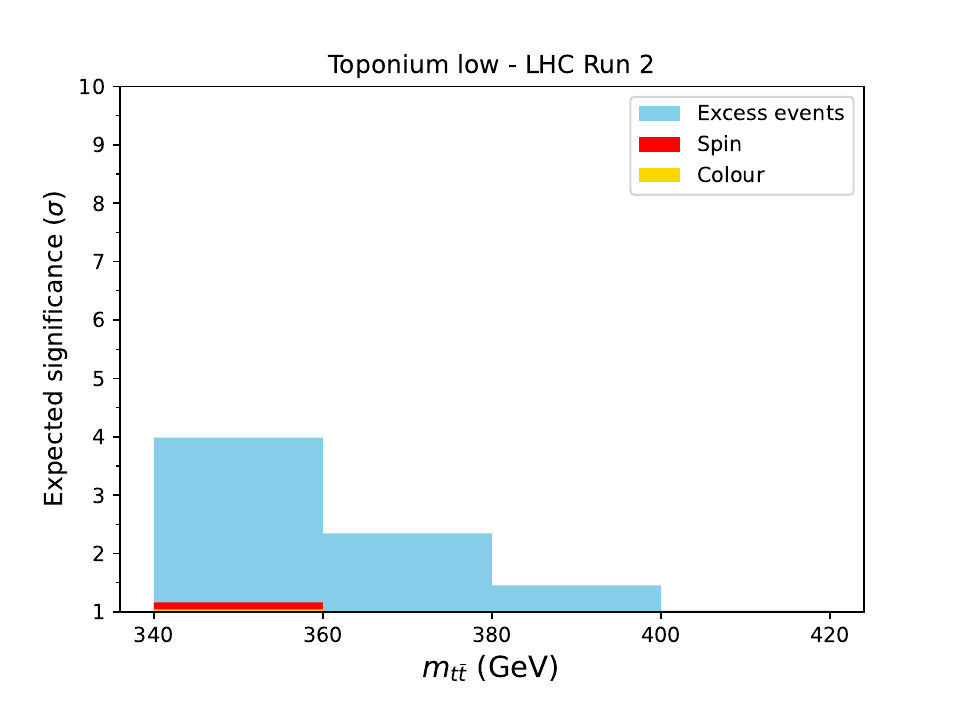} &
\includegraphics[width=8.4cm,clip=]{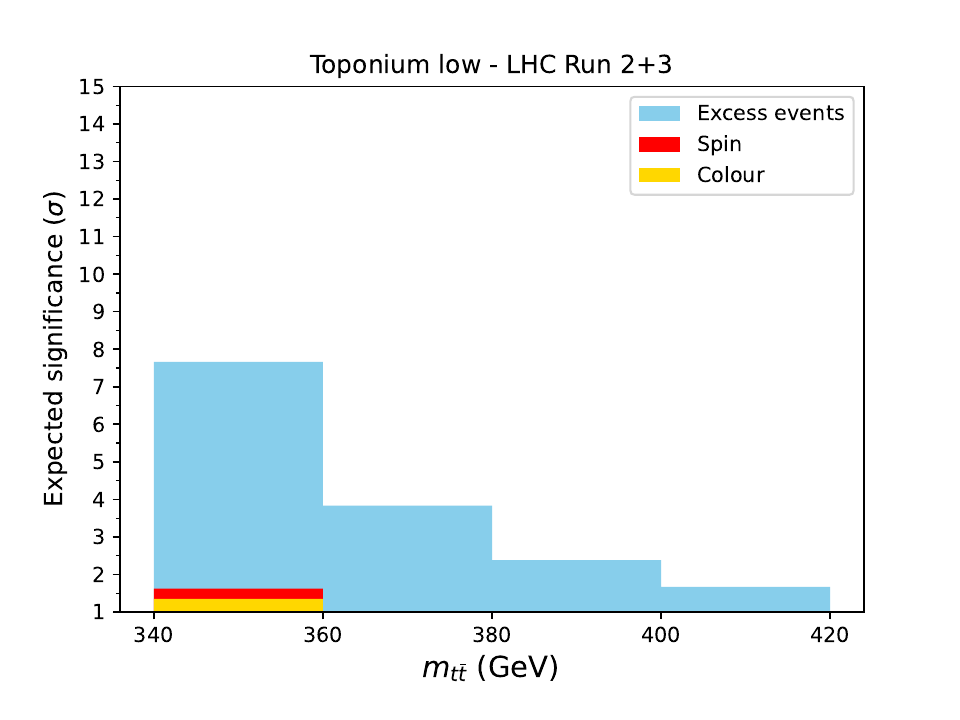} \\
\includegraphics[width=8.4cm,clip=]{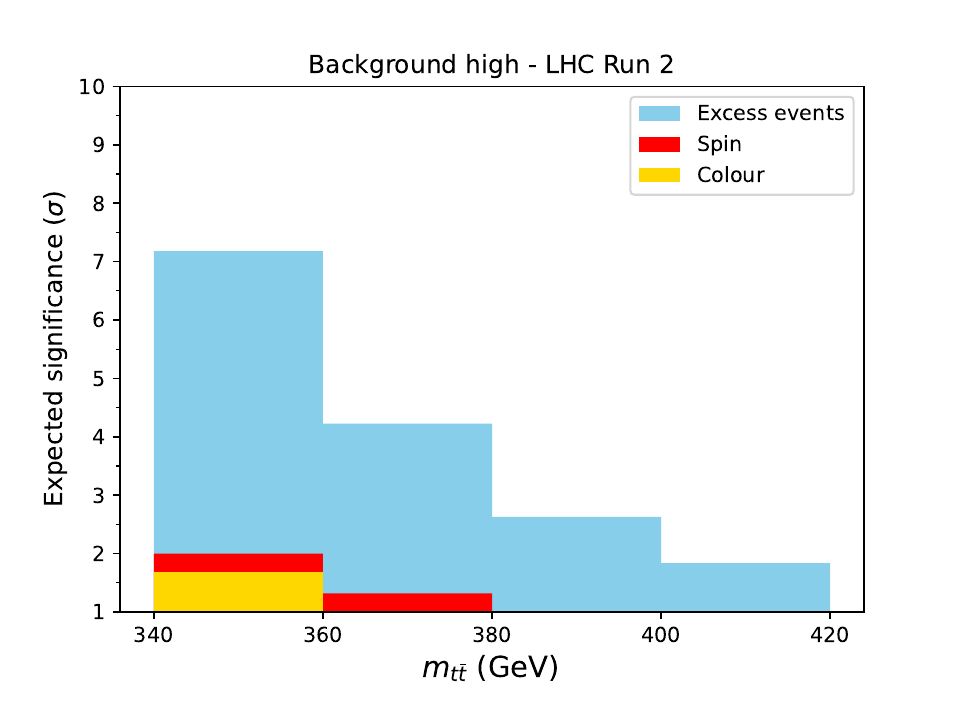} &
\includegraphics[width=8.4cm,clip=]{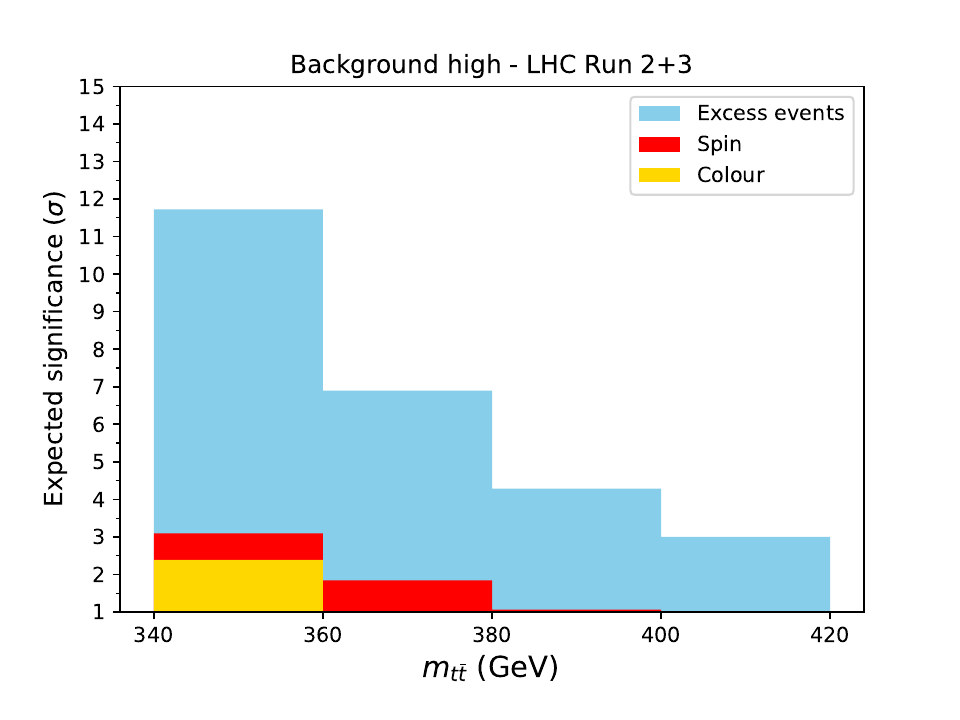}
\end{tabular}
\end{center}
\caption{Expected statistical significance for the toponium signal with Run 2 and Run 2$+$3 data, for three benchmarks described in the text.} 
\label{fig:signif}
\end{figure*}

For the excess events the expected statistical significance in each bin is computed as $S/\sqrt{B}$. The expected statistical significance of the deviations in the multi-dimensional probability density functions of spin and colour variables is calculated from pseudo-experiments. In each pseudo-experiment, two samples are compared: (i) a sample consisting of $S$ signal and $B$ background events, which is taken as pseudo-data; (ii) a sample of $S+B$ background events. Both signal and random events are randomly chosen from large pools of Monte-Carlo generated events. Repeating the pseudo-experiments $10^4$ times, we obtain the probability density function (p.d.f.) for the expected significance, from which we take the mean value as expected significance. Likewise, one can perform pseudo-experiments comparing two independent samples of $S$ signal and $B$ background events, to verify the level of agreement of pseudo-data with toponium production. As an example, we show the results for spin variables in the bin $m_{t \bar t} \leq 360$ GeV in fig.~\ref{fig:Zscore}. The size of the samples corresponds to Run 2$+$3 luminosity, in the baseline benchmark. The red distribution shows the disagreement of pseudo-data with the $b \bar b WW$ hypothesis, while the green distribution exhibits the agreement with the $b \bar b WW+\eta_t$ hypothesis. Of course, the latter depends on the toponium cross section assumed, but this could be fixed from data by the excess events observed.

The statistical significance of the deviations with respect to the $b \bar b WW$ hypothesis is presented in Fig.~\ref{fig:signif} for luminosities corresponding to LHC Run 2 and Run 2$+$3, and the three benchmarks mentioned:
\begin{itemize}
\item[(i)] Baseline: $\sigma(\eta_t) = 6.43$ pb, $k = 1.1$;
\item[(ii)] `toponium low', with $\sigma(\eta_t) = 3.06$ pb, $k = 1.1$;
\item[(iii)] `background high', with $\sigma(\eta_t) = 6.43$ pb, $k = 1.5$.
\end{itemize}
For each bin of $m_{t \bar t}$ the height of the bars indicate the expected statistical significance of the toponium signal. We discard significances lower than $1\sigma$. As it can be readily observed, the toponium signal produces an excess of events near threshold whose spin and colour properties are not consistent with continuum $t \bar t$ production but would agree with the production of a colour-singlet scalar. This excess could be visible, and identified as toponium, provided the experimental and modeling uncertainties are under control, as discussed in the next section.

\section{Discovery strategies}

The discovery and characterisation of toponium is a formidable task that requires a very accurate theoretical modeling. The dilepton decay mode is particularly clean, with quite small backgrounds. Therefore, one expects that the main source of systematic uncertainties will be the theoretical modeling, in addition to experimental uncertainties. 

The toponium signal leads to localised deviations close to the $t \bar t$ threshold, as it can be seen in Fig.~\ref{fig:signif}. These deviations decrease quickly, and for $m_{t \bar t} \geq 400$ GeV the toponium effect on spin and colour observables is below $1\sigma$ (statistical) even for Run 2$+$3 data. Then, it is conceivable that experiments could use a `control' region $m_{t \bar t} \geq 400$ GeV to tune the theoretical predictions for the relevant observables, so that their uncertainty in the `measurement' region $m_{t \bar t} \leq 360$ GeV can be reduced and the toponium signal is visible. We note that these are detector-level values, and the difference with the parton-level $m_{t \bar t}$ is in the $10-20$ GeV ballpark, see fig.~\ref{fig:reco3}. Therefore, the region $m_{t \bar t} \leq 360$ GeV at the detector level is not much stricter than the ones used by the ATLAS and CMS Collaborations in the entanglement measurements~\cite{ATLAS:2019zrq,CMS:2019nrx}. 

The identification of the event excess as toponium requires to verify that the features of the excess are (i) not compatible with $b \bar b WW$ production; (ii) compatible with $b \bar b WW+\eta_t$ production. In order to verify the latter, which requires as input the toponium cross section, one can directly use data, i.e. the size of the event excess. Theoretical predictions of the toponium cross section have a large uncertainty, as evidenced by the different cross sections
$\sigma(\eta_t) = 6.43$ pb, $\sigma(\eta_t) = 3.06$ pb obtained from Refs.~\cite{Sumino:2010bv} and ~\cite{Kiyo:2008bv}, respectively. And, in any case, the toponium cross section extracted from the dilepton and semileptonic decay modes would have to be in agreement with theoretical expectations. 

The theoretical modeling of the observables used to characterise colour seems especially difficult. However, we point out that a multi-dimensional NN discriminant based on $\tau_n^{(\beta)}$ variables may be more resilient against mismodeling than the input variables themselves. An example of this nice behaviour was found in Ref.~\cite{Aguilar-Saavedra:2022ejy} for jet taggers. While the subjettiness variables used in the NN exhibit differences depending on the hadronisation scheme, the NN discriminant turns out to be rather insensitive. 

In summary, the experimental discovery of toponium and characterisation of its properties seems a quite difficult endeavour, yet not impossible, as we have argued. Improved modeling and data-driven calibration of theoretical predictions might render visible the expected deviations in the number of events, spin, and colour observables that we have pointed out in this work.

\section*{Acknowledgements}
I gratefully acknowledge many useful discussions with M.L. Mangano and help with toponium modelling, and the CERN Theory Department for hospitality during the realisation of this work. I also thank C. Severi for correspondence regarding Ref.~\cite{Maltoni:2024csn}.
This work has been supported by the Spanish Research Agency (Agencia Estatal de Investigaci\'on) through projects PID2022-142545NB-C21 and CEX2020-001007-S funded by MCIN/AEI/10.13039/501100011033, and by Funda\c{c}{\~a}o para a Ci{\^e}ncia e a Tecnologia (FCT, Portugal) through the project CERN/FIS-PAR/0019/2021.

\appendix
\label{sec:a}
\section{Analysis of the semileptonic decay channel}

Despite the larger branching ratio, the semileptonic final state $b \bar b \ell \nu q \bar q'$ is not competitive with the dilepton one for toponium characterisation. We focus here on the lowest invariant mass bin $m_{t \bar t} \leq 360$ GeV. We generate toponium ($10^6$ events) and $b \bar b WW$ ($6 \times 10^6$ events) in the semileptonic decay mode, with a generator-level upper cut on the total invariant mass $M \leq 400$ GeV. Events are processed with {\scshape Pythia} and {\scshape Delphes} as described for the dilepton final state. For event selection we require (see e.g. Ref.~\cite{CMS:2022voq}):
\begin{itemize}
\item One charged lepton $\ell$ with $|\eta| \leq 2.4$ and $p_T \geq 30$ GeV.
\item Two $b$-tagged jets with $|\eta| \leq 2.4$ and $p_T \geq 30$ GeV.
\item At least two untagged jets with $|\eta| \leq 2.4$ and $p_T \geq 30$ GeV.
\end{itemize}
The reconstruction of the $W$ and top quark momenta is done as follows. The momentum of the $W$ boson decaying hadronically is reconstructed as $p_{W_h} = p_{j_1} + p_{j_2}$,
with $j_1$, $j_2$ the two non-tagged jets, among the three with largest $p_T$, which have invariant mass closest to $M_W$. The neutrino momentum is reconstructed defining $(p_\nu)_{x,y} = (\met)_{x,y}$, and solving he quadratic equation $(p_\ell + p_\nu)^2 = M_W^2$ for $(p_\nu)_z$. (If the equation does not have real solutions we set the discriminant to zero.) The momentum of the $W$ boson decaying leptonically is then $p_{W_l} = p_\ell + p_\nu$, and the momenta of the two top quarks
\begin{align}
& p_{t_l} = p_{W_l} + p_{b_1} \,, \notag \\
& p_{t_h} = p_{W_h} + p_{b_2} \,.
\label{ec:recsemi}
\end{align}
In general there are two solutions for $(p_\nu)_z$ and two possible pairings in (\ref{ec:recsemi}). We chose the ones that minimise the quantity
\begin{equation}
\chi^2 = (m_{t_l} - m_t)^2 + (m_{t_h} - m_t)^2 \,,
\end{equation}
with $m_{t_l} = \sqrt{p_{t_l}^2}$, $m_{t_h} = \sqrt{p_{t_h}^2}$ the invariant masses of the leptonically and hadronically-decaying top quark, respectively. Spin variables are defined as outlined in section~\ref{sec:3.1}, but using for the  hadronic top decay the optimal polarimeter introduced in Ref.~\cite{Tweedie:2014yda}. Colour observables are defined as in section~\ref{sec:3.2}.

The expected number of events with $m_{t \bar t} \leq 360$ GeV for LHC Run 2 is 95830 for $b \bar bWW$ and 4150 for toponium, i.e. around 2.5 times larger than in the dilepton decay mode, c.f. table~\ref{tab:SB}. Note that the ratio $S/B = 0.041$ is quite close to the one in the dilepton channel --- leaving aside the fact that other non-$t \bar t$ backgrounds are larger in this channel than in the dilepton mode. Still, the sensitivity to toponium using spin and colour observables is much smaller, $0.8\sigma$ and $1.5\sigma$ respectively. This is quite as expected. For spin observables the use of the hadronic polarimeter, with smaller spin analysing power than the charged lepton, leads to smaller differences in the distributions. In addition, near threshold where the jets are not very energetic, it is found that often one of the two untagged jets used to reconstruct the hadronically-decaying $W$ boson actually does not correspond to the $W$ decay --- which spoils the rest-frame angular distributions. (A quality cut on the reconstructed hadronic $W$ mass does not improve the sensitivity.) This smaller sensitivity to toponium is also in agreement with the smaller entanglement significance found by the CMS Collaboration in the semileptonic mode near threshold~\cite{CMS:2024vqh}. For colour observables, the smaller sensitivity can be attributed to the additional hadronic activity that washes out the differences due to the colour connection between the top quarks.


\begin{thebibliography}{99}

\bibitem{ATLAS:2023fsd}
G.~Aad \textit{et al.} [ATLAS],
``Observation of quantum entanglement in top-quark pairs using the ATLAS detector,''
[arXiv:2311.07288 [hep-ex]].

\bibitem{CMS:2024pts}
 [CMS],
``Observation of quantum entanglement in top quark pair production in proton-proton collisions at $\sqrt{s}$ = 13 TeV,''
[arXiv:2406.03976 [hep-ex]].

\bibitem{Fadin:1990wx}
V.~S.~Fadin, V.~A.~Khoze and T.~Sjostrand,
Z. Phys. C \textbf{48}, 613-622 (1990)

\bibitem{Hagiwara:2008df}
K.~Hagiwara, Y.~Sumino and H.~Yokoya,
Phys. Lett. B \textbf{666}, 71-76 (2008)
[arXiv:0804.1014 [hep-ph]].

\bibitem{Kiyo:2008bv}
Y.~Kiyo, J.~H.~Kuhn, S.~Moch, M.~Steinhauser and P.~Uwer,
``Top-quark pair production near threshold at LHC,''
Eur. Phys. J. C \textbf{60}, 375-386 (2009)
[arXiv:0812.0919 [hep-ph]].

\bibitem{Sumino:2010bv}
Y.~Sumino and H.~Yokoya,
``Bound-state effects on kinematical distributions of top quarks at hadron colliders,''
JHEP \textbf{09}, 034 (2010)
[erratum: JHEP \textbf{06}, 037 (2016)]
[arXiv:1007.0075 [hep-ph]].

\bibitem{Fadin:1987wz}
V.~S.~Fadin and V.~A.~Khoze,
JETP Lett. \textbf{46}, 525-529 (1987)

\bibitem{Strassler:1990nw}
M.~J.~Strassler and M.~E.~Peskin,
``The Heavy top quark threshold: QCD and the Higgs,''
Phys. Rev. D \textbf{43}, 1500-1514 (1991)

\bibitem{Sumino:1992ai}
Y.~Sumino, K.~Fujii, K.~Hagiwara, H.~Murayama and C.~K.~Ng,
``Top quark pair production near threshold,''
Phys. Rev. D \textbf{47}, 56-81 (1993)




\bibitem{Fuks:2021xje}
B.~Fuks, K.~Hagiwara, K.~Ma and Y.~J.~Zheng,
``Signatures of toponium formation in LHC run 2 data,''
Phys. Rev. D \textbf{104}, no.3, 034023 (2021)
[arXiv:2102.11281 [hep-ph]].




\bibitem{ATLAS:2019zrq}
M.~Aaboud \textit{et al.} [ATLAS],
``Measurements of top-quark pair spin correlations in the $e\mu$ channel at $\sqrt{s} = 13$ TeV using $pp$ collisions in the ATLAS detector,''
Eur. Phys. J. C \textbf{80}, no.8, 754 (2020)
[arXiv:1903.07570 [hep-ex]].

\bibitem{CMS:2019nrx}
A.~M.~Sirunyan \textit{et al.} [CMS],
``Measurement of the top quark polarization and $\mathrm{t\bar{t}}$ spin correlations using dilepton final states in proton-proton collisions at $\sqrt{s} =$ 13 TeV,''
Phys. Rev. D \textbf{100}, no.7, 072002 (2019)
[arXiv:1907.03729 [hep-ex]].

\bibitem{Behring:2019iiv}
A.~Behring, M.~Czakon, A.~Mitov, A.~S.~Papanastasiou and R.~Poncelet,
``Higher order corrections to spin correlations in top quark pair production at the LHC,''
Phys. Rev. Lett. \textbf{123}, no.8, 082001 (2019)
[arXiv:1901.05407 [hep-ph]].

\bibitem{topLHCwg}
See https://twiki.cern.ch/twiki/bin/view/LHCPhysics/ LHCTopWGSummaryPlots


\bibitem{Alwall:2014hca}
J.~Alwall, R.~Frederix, S.~Frixione, V.~Hirschi, F.~Maltoni, O.~Mattelaer, H.~S.~Shao, T.~Stelzer, P.~Torrielli and M.~Zaro,
``The automated computation of tree-level and next-to-leading order differential cross sections, and their matching to parton shower simulations,''
JHEP \textbf{07}, 079 (2014)
[arXiv:1405.0301 [hep-ph]].

\bibitem{NNPDF:2017mvq}
R.~D.~Ball \textit{et al.} [NNPDF],
``Parton distributions from high-precision collider data,''
Eur. Phys. J. C \textbf{77}, no.10, 663 (2017)
[arXiv:1706.00428 [hep-ph]].

\bibitem{Maltoni:2024csn}
F.~Maltoni, C.~Severi, S.~Tentori and E.~Vryonidou,
``Quantum tops at circular lepton colliders,''
[arXiv:2404.08049 [hep-ph]].

\bibitem{Sjostrand:2007gs}
T.~Sjostrand, S.~Mrenna and P.~Z.~Skands,
``A Brief Introduction to PYTHIA 8.1,''
Comput. Phys. Commun. \textbf{178}, 852-867 (2008)
[arXiv:0710.3820 [hep-ph]].

\bibitem{deFavereau:2013fsa}
J.~de Favereau \textit{et al.},
``DELPHES 3, A modular framework for fast simulation of a generic collider experiment,''
JHEP \textbf{02}, 057 (2014)
[arXiv:1307.6346 [hep-ex]].

\bibitem{Cacciari:2011ma}
M.~Cacciari, G.~P.~Salam and G.~Soyez,
``FastJet User Manual,''
Eur. Phys. J. C \textbf{72}, 1896 (2012)
[arXiv:1111.6097 [hep-ph]].

\bibitem{Cacciari:2008gp}
M.~Cacciari, G.~P.~Salam and G.~Soyez,
``The anti-$k_t$ jet clustering algorithm,''
JHEP \textbf{04}, 063 (2008)
[arXiv:0802.1189 [hep-ph]].

\bibitem{CMS:2012feb}
S.~Chatrchyan \textit{et al.} [CMS],
``Identification of b-Quark Jets with the CMS Experiment,''
JINST \textbf{8}, P04013 (2013)
[arXiv:1211.4462 [hep-ex]].







\bibitem{Aguilar-Saavedra:2021ngj}
J.~A.~Aguilar-Saavedra, M.~C.~N.~Fiolhais, P.~Mart\'\i{}n-Ramiro, J.~M.~Moreno and A.~Onofre,
``A template method to measure the $t {\bar{t}}$ polarisation,''
Eur. Phys. J. C \textbf{82}, no.2, 134 (2022)
[arXiv:2111.10394 [hep-ph]].

\bibitem{CMS:2016lmd}
V.~Khachatryan \textit{et al.} [CMS],
``Jet energy scale and resolution in the CMS experiment in pp collisions at 8 TeV,''
JINST \textbf{12}, no.02, P02014 (2017)
[arXiv:1607.03663 [hep-ex]].





\bibitem{Williams:2010vh}
M.~Williams,
``How good are your fits? Unbinned multivariate goodness-of-fit tests in high energy physics,''
JINST \textbf{5}, P09004 (2010)
[arXiv:1006.3019 [hep-ex]].

\bibitem{K}
A.~N. Kolmogorov, ``Sulla determinazione empirica di una legge di distribuzione,''
 Giornale dell'Istituto Italiano degli Attuari, 4, 83-91 (1933).

\bibitem{S}
N.~V. Smirnov, ''Table for Estimating the Goodness of Fit of Empirical Distributions,'' The Annals of Mathematical Statistics, 19(2), 279-281 (1948)

\bibitem{Grosso:2023scl}
G.~Grosso, M.~Letizia, M.~Pierini and A.~Wulzer,
``Goodness of fit by Neyman-Pearson testing,''
SciPost Phys. \textbf{16}, 123 (2024)
[arXiv:2305.14137 [hep-ph]].


\bibitem{Adam}
D.~P. Kingma and J. Ba,
arXiv:1412.6980 [cs.LG].

\bibitem{keras}
F. Chollet, Keras: Deep Learning for Python (2015), {\tt https://github.com/fchollet/keras}

\bibitem{tensorflow}
M. Abadi et. al., TensorFlow: Large-Scale Machine Learning on Heterogeneous Systems
(2015), {\tt http://tensorflow.org/}










\bibitem{Bernreuther:2015yna}
W.~Bernreuther, D.~Heisler and Z.~G.~Si,
``A set of top quark spin correlation and polarization observables for the LHC: Standard Model predictions and new physics contributions,''
JHEP \textbf{12}, 026 (2015)
[arXiv:1508.05271 [hep-ph]].


\bibitem{Thaler:2010tr}
J.~Thaler and K.~Van Tilburg,
``Identifying Boosted Objects with N-subjettiness,''
JHEP \textbf{03}, 015 (2011)
[arXiv:1011.2268 [hep-ph]].

\bibitem{Thaler:2011gf}
J.~Thaler and K.~Van Tilburg,
``Maximizing Boosted Top Identification by Minimizing N-subjettiness,''
JHEP \textbf{02}, 093 (2012)
[arXiv:1108.2701 [hep-ph]].

\bibitem{Datta:2017rhs}
K.~Datta and A.~Larkoski,
``How Much Information is in a Jet?,''
JHEP \textbf{06}, 073 (2017)
[arXiv:1704.08249 [hep-ph]].

\bibitem{Banfi:2010pa}
A.~Banfi, M.~Dasgupta, K.~Khelifa-Kerfa and S.~Marzani,
``Non-global logarithms and jet algorithms in high-pT jet shapes,''
JHEP \textbf{08}, 064 (2010)
[arXiv:1004.3483 [hep-ph]].

\bibitem{Gallicchio:2010sw}
J.~Gallicchio and M.~D.~Schwartz,
``Seeing in Color: Jet Superstructure,''
Phys. Rev. Lett. \textbf{105}, 022001 (2010)
[arXiv:1001.5027 [hep-ph]].

\bibitem{D0:2011lzz}
V.~M.~Abazov \textit{et al.} [D0],
``Measurement of Color Flow in $\mathbf{t\bar{t}}$ Events from $\mathbf{p\bar{p}}$ Collisions at $\mathbf{\sqrt{s}=1.96}$ TeV,''
Phys. Rev. D \textbf{83}, 092002 (2011)
[arXiv:1101.0648 [hep-ex]].

\bibitem{ATLAS:2015ytt}
G.~Aad \textit{et al.} [ATLAS],
``Measurement of colour flow with the jet pull angle in $t\bar{t}$ events using the ATLAS detector at $\sqrt{s}=8$ TeV,''
Phys. Lett. B \textbf{750}, 475-493 (2015)
[arXiv:1506.05629 [hep-ex]].





\bibitem{Czakon:2013goa}
M.~Czakon, P.~Fiedler and A.~Mitov,
``Total Top-Quark Pair-Production Cross Section at Hadron Colliders
Through $O(\alpha^4_S)$,''
Phys. Rev. Lett. \textbf{110}, 252004 (2013)
[arXiv:1303.6254 [hep-ph]].


\bibitem{CMS:2022voq}
A.~Tumasyan \textit{et al.} [CMS],
``Search for CP violation using $ \textrm{t}\overline{\textrm{t}} $ events in the lepton+jets channel in pp collisions at $ \sqrt{s} $ = 13 TeV,''
JHEP \textbf{06}, 081 (2023)
[arXiv:2205.02314 [hep-ex]].

\bibitem{Tweedie:2014yda}
B.~Tweedie,
``Better Hadronic Top Quark Polarimetry,''
Phys. Rev. D \textbf{90}, no.9, 094010 (2014)
[arXiv:1401.3021 [hep-ph]].



\bibitem{Aguilar-Saavedra:2022ejy}
J.~A.~Aguilar-Saavedra,
``Taming modeling uncertainties with mass unspecific supervised tagging,''
Eur. Phys. J. C \textbf{82}, no.3, 270 (2022)
[arXiv:2201.11143 [hep-ph]].

\bibitem{CMS:2024vqh}
CMS Collaboration,
``Measurements of polarization, spin correlations, and entanglement in top quark pairs using lepton+jets events from pp collisions at $\sqrt{s}=13~\mathrm{TeV}$,''
CMS-PAS-TOP-23-007.

\end{thebibliography}
\end{document}